\documentclass[aps,twocolumn,superscriptaddress,showpacs,floatfix]{revtex4}
\usepackage{dcolumn}
\usepackage{graphicx}
\usepackage{amsfonts}
\usepackage{amsmath,bm,amssymb}

\begin{document}

\title{Competition between Kondo and RKKY exchange couplings in Pu$_{1-x}$Am$%
_{x}$ alloys}
\author{Myung Joon Han}
\affiliation{Department of Physics, University of California, Davis, California 95616, USA}
\author{Xiangang Wan}
\affiliation{Department of Physics, University of California, Davis, California 95616, USA}
\affiliation{National Laboratory of Solid State Microstructures and Department of
Physics, Nanjing University, Nanjing 210093, China}
\author{Sergej Y. Savrasov}
\affiliation{Department of Physics, University of California, Davis, California 95616, USA}
\date{\today }

\begin{abstract}
To clarify the role of the Kondo effect in screening local magnetic moments
of Plutonium 5f--electrons as well as its competition to the RKKY
interactions we use a combination of density functional theory with static
Hartree Fock and dynamic Hubbard 1 approximations to calculate the strength
of both the Kondo exchange, $J_{\text{K}}$, and of the RKKY exchange, $J_{%
\text{RKKY}}$, couplings for Pu$_{1-x}$Am$_{x}$ system as a function of $x$.
We find that $J_{\text{K}}$ increases despite the atomic volume gets larger
with the Am doping due to unexpected enhancement of hybridization between $f$
and conduction electrons in the vicinity of the Fermi level. At the same
time, the RKKY\ exchange is shown to reduce smoothly with increasing $x$.
Our results imply that the Kondo effect should be robust against the
increase in interatomic spacing of this alloy.
\end{abstract}

\maketitle


Magnetic behavior of metallic Plutonium is a topic of hot debates in the
current literature \cite{Lashley}. Naively one expects that Pu f--shell
filled with 5 electrons carries a total (spin+orbital) momentum which
depending on various spin--orbit coupling schemes and crystal field effects
should result in some non--zero values of magnetic moment. This has been
confirmed by many state--of--the--art electronic structure calculations \cite%
{Solovyev,Eriksson,Singh,Soderlind} based on density functional theory (DFT)
in its local density and generalized gradient approximations (LDA and GGA) 
\cite{LDA} and by methods such as LDA+U \cite{Pu-LDA+U1,Pu-LDA+U2} allowing
to incorporate effects of on--site correlation energy $U$ using static
Hartree--Fock type approximations for the f--electron self--energies \cite%
{LDA+U}. Experimentally, however, none of the six Pu crystallographic
allotropes shows local moment formation: their spin susceptibilities are
small, temperature independent and display Pauli--like behavior \cite%
{Sandenaw}; specific heat measurements indicate absence of magnetic entropy 
\cite{SHeat}; muon experiments did not detect the moment with the accuracy
of 10$^{-3}$ $\mu _{B.}$\cite{muon}; combinations of neutron elastic \cite%
{Lawson} and inelastic \cite{INS} scattering data show no convincing
evidence in existence of magnetic moments either.

The apparent discrepancy between theory and experiment can be attributed to
miscounting the number of f--electrons and can assume the true configuration
to be f$^{6}$ as it was pointed using a variant of the LDA+U calculation 
\cite{ShickPu} with a different prescription to determine the position of
the f--band (so called choice of double counting potential). However, this
would produce a completely inert f--shell and no large effective masses for
the Fermi surface electrons which contradicts sharply with specific heat
data exhibiting an enhanced Sommerfeld coefficient \cite{Lashley}.
Many--body based calculations \cite{Shim} based on a combination of density
functional and dynamical mean field \cite{DMFT} theories (so called LDA+DMFT 
\cite{RMP}) have recently highlighted another effect: the Kondo coupling of
Pu f electrons which fluctuate between f$^{5}$ or f$^{6}$ atomic like states
with the conduction band made of $sd$ electrons producing the Kondo singlet.

To understand this puzzle, one can try to increase Pu atomic volume in order
to reduce the effect of hybridization and thus to decrease the value of the
Kondo coupling $J_{\text{K}}.$ As a result, if the Kondo screening mechanism
is in play one expects that at some critical interatomic distance the local
moment would eventually show up. Indeed most recent studies\cite{Marianetti}
of stretched Plutonium lattice based on the LDA+DMFT calculations with
Continuous--Time Quantum Monte Carlo (CT--QMC)\ method for the solution of
impurity problem \cite{CTQMC} have detected that at volumes of the order of
30\% larger than the volume of the fcc $\delta -$phase the temperature
behavior of spin susceptibilities turns from Pauli-like to Curie--like. To
simulate similar stretch experimentally, Pu$_{1-x}$Am$_{x}$ alloys have been
made, which, however, on the basis of magnetic susceptibility, electrical
resistivity, and photoelectron spectroscopy studies \cite%
{Gouder,Javorsky,Baclet}, indicated that the character of the 5f states does
not vary with the Am doping. This came into a contradiction with the DFT
based study of this system\cite{Landa}.

We thus see that deducing the behavior of the Kondo exchange energy as a
function of $x$ is an interesting theoretical problem which may shed a light
on the description of the recent experimental results in the Pu--Am alloy.
As a minimal model applicable for this description is the model of the Kondo
lattice, it is also interesting to understand an approximate location of
this system in the Doniach phase diagram \cite{Doniach} where the
competition between Kondo and RKKY interactions may lead to exciting
phenomena of quantum criticality and exotic superconductivity. This, for
example, is seen in a different class of Pu--based 115 materials such as
PuCoGa$_{5}$ and PuRhGa$_{5}$ \cite{115}.

In the present work, both the Kondo coupling strength, $J_{\text{K}}$, and
the hypothetical magnetic interaction strength, $J_{\text{RKKY}}$, are
calculated and compared against each other as a function of $x$ using
supercell based electronic structure framework where the f--electrons are
treated as localized. To deduce $J_{\text{K}}$ we measure the hybridization
between the $f$ level and the Fermi surface $sd$ electrons. To deduce $J_{%
\text{RKKY}}$ we use a newly developed linear response approach \cite{Wan}
based on magnetic force theorem\cite{J-RKKY} . The f--electron
self--energies are approximated by their atomic Hartree--Fock values using
the LDA+U method but the results are checked against the LDA+DMFT
calculations with the self--energies extracted by exact diagonalizing
many--body atomic Hamiltonians \cite{Hub1}. The main finding of our work is
an unexpected increase of $J_{\text{K}}$ with Am doping due to a particular
behavior of the hybridization function in the vicinity of the Fermi level
and simultaneous decrease of $J_{\text{RKKY}}.$ The latter is expected as we
scale up the interatomic distances of the lattice. Our calculation shows
that $J_{\text{K}}$ always remains larger than $J_{\text{RKKY}}$ assuming
that the Kondo effect is robust against the increase in atomic volume of Pu$%
_{1-x}$Am$_{x}$ which upon doping transforms from the Kondo lattice to the
diluted impurity limit . It also places this system into the heavy fermion
region of the Doniach phase diagram away from the quantum critical behavior.

Our calculations are performed using the full potential
linearized--muffin--tin--orbital (LMTO) method including relativistic effect
of spin--orbit coupling \cite{Savrasov-1996}. An effective $U=4.5$ eV
describing the on--site Coulomb repulsion among the $5f$ electrons is used
while the other Slater integrals ($F^{(2)}$, $F^{(4)}$ and $F^{(6)}$) are
computed from atomic physics, and are subsequently rescaled to $80\%$ of
their values to account for the effect of screening \cite{18}. Those numbers
are known to give a reasonable description of the electronic structure for
both Pu and Pu--Am system \cite{Savrasov-2000,Shick-2006,Savrasov-2006}. To
simulate the effect of alloying in our study we used supercells with 4 atoms
corresponding to $x=0$,$\frac{1}{4}$,$\frac{1}{2},\frac{3}{4},1$ in Pu$%
_{1-x} $Am$_{x}$ phase diagram, but the results of $x\geq \frac{3}{4}$
cannot be taken into account for $J_{RKKY\ }$because it corresponds to a
ferromagnetic (FM) phase ($x=\frac{3}{4}$) and dilute impurity limit ($%
x\rightarrow 1$). We have also utilized experimental lattice parameters for
various $x$\cite{Ellinger}$.$Our calculated electronic structures are found
to be consistent with the previous studies \cite{Baclet,Shick-2006}: the 5f
states of Am are well localized and reside in their $5f^{6}$ configuration,
corresponding to the filled 5/2 shell; Pu $5f$ states are not fully occupied
and found in configuration with 5.4 electrons; Around the Fermi energy the $%
sd$ conduction electrons prevail through which Pu local moments can interact
with each other via the RKKY\ mechanism and get screened via the Kondo
effect.

In order to estimate the strength of the Kondo coupling we calculate the
hybridization function between the $5f$ and the conduction states, $\Delta
_{\alpha \beta }(\omega )$, which is generally expressed via the local Green
function for the f--electrons as follows \cite{RMP}%
\begin{equation}
\Delta _{\alpha \beta }(\omega )=\omega o_{\alpha \beta }-\epsilon _{\alpha
\beta }-G_{\alpha \beta }^{-1}(\omega )+\Sigma _{\alpha \beta }(\omega )
\label{D}
\end{equation}%
where%
\begin{equation}
G_{\alpha \beta }(\omega )=\sum_{\mathbf{k}}(\omega \hat{O}^{\mathbf{k}}-%
\hat{H}^{\mathbf{k}}-\Delta \hat{\Sigma}(\omega )]_{\alpha \beta }^{-1}
\label{G}
\end{equation}%
Here $\hat{H}_{\mathbf{k}},\hat{O}_{\mathbf{k}}$, are single particle LDA
Hamiltonian and overlap matrix written in a general non--orthogonal LMTO
orbital basis while $\Delta \hat{\Sigma}(\omega )$ is the self--energy
correction appeared in the f--electron block only with the double counting
potential subtracted. The matrix of impurity levels $\epsilon _{\alpha \beta
}$ and the average overlap integrals $o_{\alpha \beta }$ can be found
straightforwardly from $\hat{H}^{\mathbf{k}},\hat{O}^{\mathbf{k}}$ \cite{RMP}%
. Here we utilize the static Hartree Fock approximation for the self--energy
by forcing antiferromagnetically ordered state. This is done to keep the
same level of accuracy with our subsequent evaluations of the RKKY\
interactions. We benchmarked these calculations against the Hubbard 1
approximation \cite{Hub1} assuming magnetically disordered solutions and
find similar results. Both methods do not assume DMFT\ self--consistency
with respect to the hybridization function and are reduced to the
self--consistent determination of charge densities similar to the Kohn--Sham
procedure in DFT. The imaginary part of the hybridization function taken at
zero frequency determines the strength of the Kondo exchange according to a
simple estimate\cite{SchriefferWolf} 
\begin{equation}
J_{\text{K}}=\frac{Tr\{{Im}\Delta (0)\}}{\pi N_{d}N(0)}\frac{U}{%
\epsilon _{f}(\epsilon _{f}+U)},  \label{JK}
\end{equation}%
where $N_{d}$ is the corresponding degeneracy of the model, $N(0)$ is the
density of states at the Fermi level, and $\epsilon _{f}=Tr\{\hat{\epsilon}%
\}/N_{d}.$ Thus we see that all the parameters in this expression can be
evaluated in our calculation where we find that the average position of the
impurity level $\epsilon _{f}\approx $1 eV and the total density of states $%
N(0)\approx 1.5$ states /[eV$\cdot $atom] at the Fermi level are weakly
dependent functions of Am concentration, and the trend in $J_{\text{K}}$ is
mainly determined by the behavior of ${Im}\Delta (0)$.

\begin{figure}[tbp]
\centering \includegraphics[width=6cm]{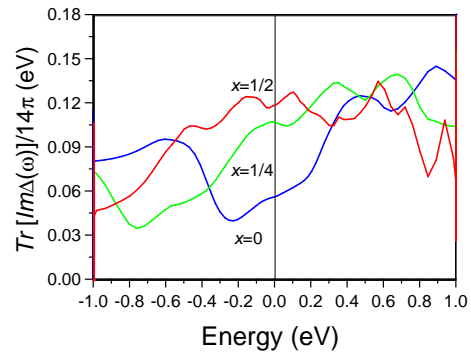}
\caption{(Color online) Calculated hyubridization function $Tr\{{Im}%
\Delta (\protect\omega )\}/N_{d}\protect\pi $ around the Fermi level for Pu$%
_{1-x}$Am$_{x}$. Lines correspond to the Am doping with $x=0,\frac{1}{4}$,
and $\frac{1}{2}$.}
\label{Delta}
\end{figure}

Figure \ref{Delta} illustrates our calculated behavior of $Tr\{{Im}%
\Delta (\omega )\}/N_{d}\pi $ for frequencies around the Fermi level and for
doping levels $x\leq \frac{1}{2}$ assuming full degeneracy $N_{d}=14$. Three
various lines correspond to the dopings with $x=0,\frac{1}{4},$ and $\frac{1%
}{2}$. The following conclusions can be derived. First, all curves look very
similar although there is an almost rigid shift of the order of 0.4 eV which
separates the calculated ${Im}\Delta (\omega )$ for various $x$.
Second, we monitor the overall trend of decreasing the hybridization between
the $f$ and conduction electrons with the Am doping as one can trace the
maximum of each plot. This is easily understood since interatomic distances
get larger. However, importantly that for all doping ratios, ${Im}%
\Delta (\omega )$ has a valley and a hilltop at the right hand side of the
valley. For pure Pu, the Fermi level is located close to the dip and it
gradually climbs up to the hilltop as $x$ increases. This results in an
unexpected increase in hybridization for $\omega =0$: $Tr\{{Im}\Delta
(0)\}/N_{d}\pi =$0.05 eV for $x=0$; but jumps to 0.10 eV for $x=\frac{1}{4}$%
. (see Table \ref{tab:J} for full compilation of the data). Accordingly, $J_{%
\text{K}}$ also increases because $N(0)$ remains approximately the same for
all $x$ values. If one sets $N_{d}=14$ in Eq. (\ref{JK})$,$ then for pure
Pu, $J_{\text{K}}$ is 340 K but becomes 660 K for $x=\frac{1}{4}$ and
further raises to 770K for $x=\frac{1}{2}$. Then $J_{K}$ decreases but is
still sufficiently large as $x$ approaches the dilute impurity limit. One
thus concludes that the Kondo screening is the robust effect upon the Am
doping which would prevent the Pu moment to appear at all $x$ .

As the description in terms of the Kondo lattice Hamiltonian may be relevant
for Pu$_{1-x}$Am$_{x}$ system, its properties should be controlled by the
competitions between the Kondo and RKKY exchange interactions which,
according to the Doniach phase diagram, depending on the precise value of $%
J_{\text{K}}$ may lead to either weakly coupled magnetically ordered local
moment state or to the Kondo screened heavy fermion state in the strong
coupling limit. It can even put the system in the vicinity of quantum
critical point where exotic superconductivity is believed to occur. In the
approximation when only a single conduction band hybridizes with the
f--level, $J_{\text{RKKY}}\ $scales simply as $J_{\text{K}}^{2}N(0)$ as seen
by using the second--order perturbation theory for the Coqblin--Schrieffer
Hamiltonian\cite{CoqlbinSchrieffer}. It may therefore be expected at first
glance that both $J_{\text{RKKY}}$ and $J_{\text{K}}$ should behave
similarly upon doping. However, in realistic situations detailed electronic
structure of the material matters as various interband transitions
contribute to exchange processes and this simple trend may be violated.

In order to estimate the strength of the magnetic interaction between
localized $5f$ states appeared while mapping the Pu sublattice onto the
Heisenberg (pseudo)spin Hamiltonian, $H=\sum J_{RR^{\prime }}%
\mathbf{S}_{R}\cdot \mathbf{S}_{R^{\prime }}$we utilize the magnetic force
theorem within a rigid spin perturbation method \cite{J-RKKY}. In this
framework, $J_{RR^{\prime }}$ is given as a second--order derivative of the
total energy induced by the rotations of magnetic moments at sites $R$ and $%
R^{^{\prime }}$ which can be found by calculating the following
spin--susceptibility--type integral: 
\begin{eqnarray}
J_{RR^{^{\prime }}}^{\alpha \beta } &=&\frac{\partial ^{2}E}{\partial \phi
_{\alpha R}\partial \phi _{\beta R^{\prime }}}  \notag \\
&=&\sum_{\mathbf{q}}\sum_{\mathbf{\text{k }}jj^{^{\prime }}}\frac{f_{\mathbf{%
k}j}-f_{\mathbf{k+q}j^{^{\prime }}}}{\epsilon _{\mathbf{k}j}-\epsilon _{%
\mathbf{k+q}j^{^{\prime }}}}\langle \mathbf{k}j|[\mathbf{\sigma }\times 
\mathbf{B}_{R}]_{\alpha }|\mathbf{k}+\mathbf{q}j^{\prime }\rangle  \notag \\
&\times &\langle \mathbf{k}+\mathbf{q}j^{\prime }|[\mathbf{\sigma }\times 
\mathbf{B}_{R^{\prime }}]_{\beta }|\mathbf{k}j\rangle e^{i\mathbf{q}\cdot (%
\mathbf{R}-\mathbf{R}^{\prime })},  \label{JR}
\end{eqnarray}%
Here $f_{\mathbf{k}j}$, $\mathbf{\sigma }$, and $\mathbf{B}_{R}$ are the
Fermi function, Pauli spin matrix, and the effective magnetic field at atom $%
R$, respectively. The latter is given by the difference in the electronic
self--energies for spin up and spin down electrons. Since the LDA+U method
is employed to recover antiferromagnetically ordered state, those become
frequency independent matrices and the evaluation of interatomic exchange
interactions is straightforward \cite{Wan}. In practical calculations using
the supercells the total number of nearest neighboring $J$'s are different
for different doping ratios, and we take average values after calculating
all possible nearest $J$'s.

We find that our calculated $J_{\text{RKKY}}$ exhibits a trend opposite to $%
J_{\text{K}}$. It decreases as $x$ increases as it is evident from Table \ref%
{tab:J}. The $J_{\text{RKKY}}$ in pure Pu ($x=0$) is smaller than $J_{\text{K%
}}$, 134 K, which is reasonable in the sense that it is set by the scale $J_{%
\text{K}}^{2}N(0)$ and that from the experimental standpoint there is no
local moment in Pu due to the Kondo screening. However, the behavior is
quite different from a simple trend that $J_{\text{RKKY}}\symbol{126}J_{%
\text{K}}^{2}N(0)$ as at $x=\frac{1}{4}$, $J_{\text{RKKY}}$ =100 K, and it
becomes 67 K at $x=\frac{1}{2}$. This must be due to interband transitions
presented in Eq.(\ref{JR}). The positive sign of $J_{\text{RKKY}}$ refers to
the AFM order by the convention in Eq.\ (\ref{JR}) and it is consistent with
the assumed AFM ground state.

\begin{table}[tbp]
\caption{The calculated $Tr\{{Im}\Delta (0)\}/14\protect\pi $, $J_{%
\text{K}}$ and $J_{\text{RKKY}}$ for various dopings of Pu$_{1-x}$Am$_{x}$
alloy. }
\label{tab:J}%
\begin{tabular}{cccc}
\hline
Am ratio & $Tr\{{Im}\Delta (0)\}/14\pi $ (eV) & $J_{\text{K}}$ (K) & $%
J_{\text{RKKY}}$ (K) \\ \hline
$x=0$ & 0.05 & 340 & 134 \\ 
$x=\frac{1}{4}$ & 0.10 & 660 & 100 \\ 
$x=\frac{1}{2}$ & 0.12 & 770 & 67 \\ 
$x=\frac{3}{4}$ & 0.10 & 660 & --- \\ 
$x\rightarrow 1$ & 0.07 & 450 & --- \\ \hline
\end{tabular}%
\end{table}

The comparison of these two quantities, $J_{\text{K}}$ and $J_{\text{RKKY}}$%
, provides us with a clear picture for the magnetic properties of Pu$_{1-x}$%
Am$_{x}$. It follows that $J_{\text{K}}$ is always larger than $J_{\text{RKKY%
}}$ up to $x=\frac{1}{2}$ which covers up the whole range of the experiments
up to now \cite{Gouder,Javorsky,Baclet}. So, if the Kondo screening works
for Pu, it should also work for the alloy. Moreover, the trend is quite
suggestive as we approach the dilute limit. While we cannot extract the
value of $J_{\text{RKKY}}$ for $x\geq \frac{3}{4}$ (for $\ x=\frac{3}{4}$%
there is only 1 Pu atom left in our supercell producing FM solution), the
overall trend for $J_{\text{RKKY}}$ to decrease is expected as the inter--Pu
distances increase. It is therefore clear that $J_{\text{RKKY}}$ would
decrease further as $x$ approaches to unity. The behavior of $J_{\text{K}}$
for large $x$ is controlled by ${Im}\Delta (0)$ as the Fermi level
reaches the vicinity of the top point of the hybridization function as seen
in Fig.~\ref{Delta}. Therefore it is beginning to decrease slightly at
values of $x\geq \frac{1}{2}$. Nevertheless, even in the dilute impurity
limit, where the conduction bands are essentially made of Am $sd$ electrons,
our calculated $J_{\text{K}}$ does not drop sharply as seen from Table \ref%
{tab:J}. All this implies that the Pu--Am system is far from the quantum
critical behavior and resides in the heavy fermion state.

There are possible sources of errors in our estimates. First, the calculated 
$J_{\text{RKKY}}$ may be overestimated by the static approximation, such as
LDA+U. This, in particular, was found in the previous studies\cite{Wan,MnO}
of transition--metal oxides where the calculated exchange interactions
depending on the level of approximation for the self--energy can be further
reduced by about 10--30 \%. Second, the use of more refined impurity solvers
and corresponding effects of the DMFT self--consistency will change our
estimated values of $J_{\text{K}},$ although this effect is not expected to
be large due to (i) generally small values in the f--electron hybridization
function, and (ii), the position of the f--level at around 1 eV which is
pretty far from the Fermi energy preventing the extreme sensitivity of the
Kondo temperature. In fact, most recent LDA+DMFT\ studies of this system
based on the CT--QMC method have confirmed these conclusions \cite{JiHoon}.

In summary, using a combination of density functional theory with
self--energy corrections for the 5f--electrons we performed the estimates of
the Kondo and RKKY exchange couplings for the whole range of dopings in Pu$%
_{1-x}$Am$_{x}$ alloy. It was found that $J_{\text{K}}$ and $J_{\text{RKKY}}$
exhibit opposite trends: the $J_{\text{K}}$ increases with $x$ which is
attributed to the details in the behavior of the hybridization function near
the Fermi level while $J_{\text{RKKY}}$ is found to decrease as interatomic
distances get larger with doping. Comparing these two values provides a
clear picture of the robust Kondo effect as the origin of non--magnetic
behavior reported in recent experiments on this system.

We are thankful to\ M.J. Fluss, K. Haule, G. Kotliar and J.H. Shim for
useful conversations. This work was supported by the NSF grants No.\
0608283, 0606498 and by the US DOE grant No.\ DE-FG52-06NA2621. X.G.W.
acknowledges support from Natural Science Foundation of China under Grant
No. 10774067, and Natural Science Foundation of Jiangsu Province through
Grant No. BK2007127.

\end{document}